\documentclass[10pt,twocolumn,letterpaper]{article}

\usepackage[utf8]{inputenc} 
\usepackage[T1]{fontenc}    

\usepackage{cvpr}
\usepackage{times}
\usepackage{epsfig}
\usepackage{graphicx}
\usepackage{amsmath}
\usepackage{amssymb}
\usepackage{bm}

\usepackage[pagebackref=true,breaklinks=true,letterpaper=true,colorlinks,bookmarks=false]{hyperref}

\cvprfinalcopy 

\def\cvprPaperID{5848} 

\ifcvprfinal\fi
\begin{document}

\title{Computationally Efficient Neural Image Compression}

\author{Nick Johnston \qquad Elad Eban \qquad \qquad  Ariel Gordon \qquad   Johannes Ballé \\
Google Research \\
{\tt\small nickj@google.com} \quad {\tt\small elade@google.com} \quad {\tt\small gariel@google.com} \quad {\tt\small jballe@google.com}}


\maketitle

\begin{abstract}
	Image compression using neural networks have reached or exceeded
	non-neural methods (such as JPEG, WebP, BPG). While these networks
	are state of the art in rate-distortion performance, computational
	feasibility of these models remains a challenge. We apply automatic network optimization techniques to reduce the computational complexity of a popular
	architecture used in neural image compression, analyze the decoder
	complexity in execution run-time and explore the trade-offs between two
	distortion metrics, rate-distortion performance and run-time performance
	to design and research more computationally efficient neural image compression. We find that our method decreases the decoder run-time requirements by over 50\%
	for a state-of-the-art neural architecture.
\end{abstract}

\section{Introduction}
In recent years, tremendous progress has been made in the field of neural image compression \cite{44844,ThShCuHu17, BaLaSi17,waveone2017,balle2018variational,NIPS2018_8275,balle2018integer,liu2019nonlocal}. Each advancement produced another additional milestone in rate-distortion performance with respect to classic image compression codecs \cite{jpeg, jpeg2000, WebP, BPG} until they were first match or exceeded in Mean-Squared Error (MSE) or Multiscale Structural Similarity(MS-SSIM)~\cite{wang2003multiscale} in \cite{balle2018variational}.

Despite this success in the field of neural image compression, neural methods have not be widely deployed. A number of explanations can be found: lack of confidence in metrics, difficulty and lead time
needed to ratify and adopt a new standard, portability of the technology or run-time requirements. 
With the exception of \cite{waveone2017}, run-time metrics have not been well documented in neural image compression literature, and there is virtually no prior work explaining in detail how the model architecture was arrived at. No research documents their method for choosing architecture. In this paper, we focus on this architecture search and how it relates to both rate--distortion and run-time.

We focus on the run-time performance of decoders in this work because of the asymmetric nature of compression. Images are generally compressed once and decompressed many times. Since an image could be viewed millions of times and on compute constrained devices, like smart phones, then high performance decoding is mandatory.

Due to the ubiquity and performance requirements around image decoders, we feel a study and understanding of where neural image compression performance currently is and how to improve the decoding performance is integral to mainstream acceptance and deployment of this rich field of research.

In this work, we:
\begin{enumerate}\itemsep1pt
  \item Analyze the complexity of a common neural architecture's decoder.
  \item Optimize Generalized Divisive Normalization (GDN, an activation function commonly used in neural image compression)~\cite{BaLaSi16} to reduce run-time at no performance loss.
  \item Apply a regularization technique to optimize decoder architecture for Computationally Efficient Neural Image Compression (CENIC) models.
  \item Analyze the trade offs of these learned architectures with respect to rate--distortion performance.
\end{enumerate}

In Section~\ref{section:previouswork} we discuss previous work related to neural image compression, performance and architecture search. We introduce the methods used in Section~\ref{section:methods} including a baseline performance of the Mean-Scale Hyperprior architecture~\cite{NIPS2018_8275} (referred to as Mean-Scale), the GDN variant, and a description of the regularization technique focused on computation reduction. In Section~\ref{section:training} we describe the training procedure for our networks along with initial rate-distortion curves for the entire loss function sweep. In Section~\ref{section:results} we analyze and discuss the performance characteristics of the newly learned architectures and how they related to the rate-distortion performance. We found that it is easier to design high-performance compute-constrained models trained for MS-SSIM than for MSE. Finally, in Section~\ref{section:conclusions} we conclude with recommendations on how to apply these methods to neural image compression networks along with offering additional insights that can be achieved through future work.

\subsection{Previous Work}
\label{section:previouswork}
Image compression using neural networks has quickly become a rich field of research in recent
years with \cite{44844}, \cite{ThShCuHu17} and \cite{BaLaSi17}. 
\cite{liu2019nonlocal} has recently achieved state of the art in rate-distortion, which is based on the
combined auto-regressive and hyperprior work introduced in \cite{NIPS2018_8275}.

While most research has focused on improving the rate-distortion curve on image compression performance,
less research has been done on designing the end-to-end neural networks to be feasible in the applications space.
\cite{waveone2017} showed not only state of the art performance in MS-SSIM, but was also one of the few works to report neural compression and decompression runtimes, although the lack of details given on implementation and architecture make it difficult to understand what architectural details contribute to better runtime performance.
\cite{balle2018integer} improved on the work of \cite{balle2018variational}  and demonstrated the importance of deterministic computation needed for cross-device
deployment, necessary for practical application of this research.

Neural architecture search \cite{Zoph_2018_CVPR, 2019evolution} and model compression and pruning techniques \cite{group-lasso,lebedev2016fast,wen2016learning,alvarez2016learning,murray2015auto,gordon2017morphnet} are playing an increasing role in the research of various computer vision and machine learning problems. However, to our knowledge, such techniques have not been utilized in the past for image compression models. We use an implementation of MorphNet \cite{gordon2017morphnet} which has already shown to optimize classification network by balancing classification tasks with compute on much deeper architectures.

\section{Mean-Scale Architecture}
\label{section:methods}
\label{section:baseline}
We are using the \emph{Mean-Scale} Hyperprior network as described in \cite{NIPS2018_8275}, but without the context model.
We omit the context model for two reasons, the first being simplicity in testing on a baseline network
that is already better than BPG~\cite{BPG} and the second being the difficulty of implementing a computationally efficient autoregressive context model over latents. Some other work uses much deeper architectures. However, we choose this shallower network for analyzing run-time characteristics, because it is already more likely to be computationally more efficient than a deeper architecture.

\begin{figure}[tb]
  \centering
  \includegraphics[width=0.90\linewidth]{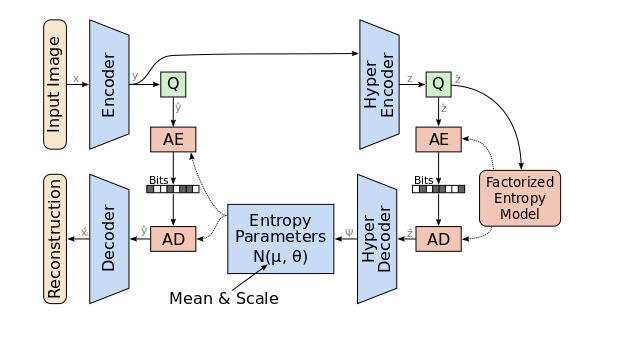}
  \caption{This is the Mean-Scale architecture we use as our baseline. 
  Figure reproduced with permission from \cite{NIPS2018_8275}. }
  \label{fig:arch}
\end{figure}

Because the Group Lasso Regularization removes channels from the network architecture, we 
purposefully start out with an over-parameterized network with the knowledge that we have less computation in the overall learned final architectures. Throughout this paper we refer to the over-parameterized networks as \emph{Larger Mean-Scale} and the learned final architectures as \emph{Computationally Efficient Neural Image Compression (CENIC)} models. Our description of the Mean-Scale architecture (Figure~\ref{fig:arch}) is described in Table~\ref{tab:baseline} and the Larger Mean-Scale architecture in Table~\ref{tab:larger}.

\begin{table}
    \begin{center}
    \scriptsize
        \begin{tabular}{c|c|c|c}
             \hline
             Encoder & Hyper Encoder & Hyper Decoder & Decoder \\
             \hline
             5x5conv,2,192 & 3x3conv,1,320 & 5x5deconv,2,320 & 5x5deconv,2,192 \\
             5x5conv,2,192 & 5x5conv,2,320 & 5x5deconv,2,320 & 5x5deconv,2,192 \\
             5x5conv,2,192 & 5x5conv,2,320 & 3x3deconv,1,320 & 5x5deconv,2,192 \\
             5x5conv,2,192 &               &                 & 5x5deconv,1,3 \\
        \end{tabular}
        \caption{Description of the Mean-Scale encoder, hyper encoder, hyper decoder and decoder network architectures. We use the notation of CxC[conv or deconv],S,F to describe either a convolutional or deconvolutional layer with a kernel of CxC, a stride of S with F output channels.}
        \label{tab:baseline}
    \end{center}
\end{table}


\begin{table}
    \begin{center}
    \scriptsize
        \begin{tabular}{c|c|c|c}
             \hline
             Encoder & Hyper Encoder & Hyper Decoder & Decoder \\
             \hline
             5x5conv,2,320 & 3x3conv,1,640 & 5x5deconv,2,640 & 5x5deconv,2,320 \\
             5x5conv,2,320 & 5x5conv,2,640 & 5x5deconv,2,640 & 5x5deconv,2,320 \\
             5x5conv,2,320 & 5x5conv,2,320 & 3x3deconv,1,320 & 5x5deconv,2,320 \\
             5x5conv,2,320 &               &                 & 5x5deconv,1,3 \\
        \end{tabular}
        \caption{Description of the Larger Mean-Scale encoder, hyper encoder, hyper decoder and decoder network architectures. The hyper encoder and decoder networks are identical for the mean and scale prediction. We use the notation of CxC[conv or deconv],S,F to describe either a convolutional or deconvolutional layer with a kernel of CxC, a stride of S with F output channels.}
        \label{tab:larger}
    \end{center}
\end{table}

\subsection{GDN activation function}
\label{section:gdn_desc}
\begin{figure}[tb]
  \centering
  \includegraphics[width=0.90\linewidth]{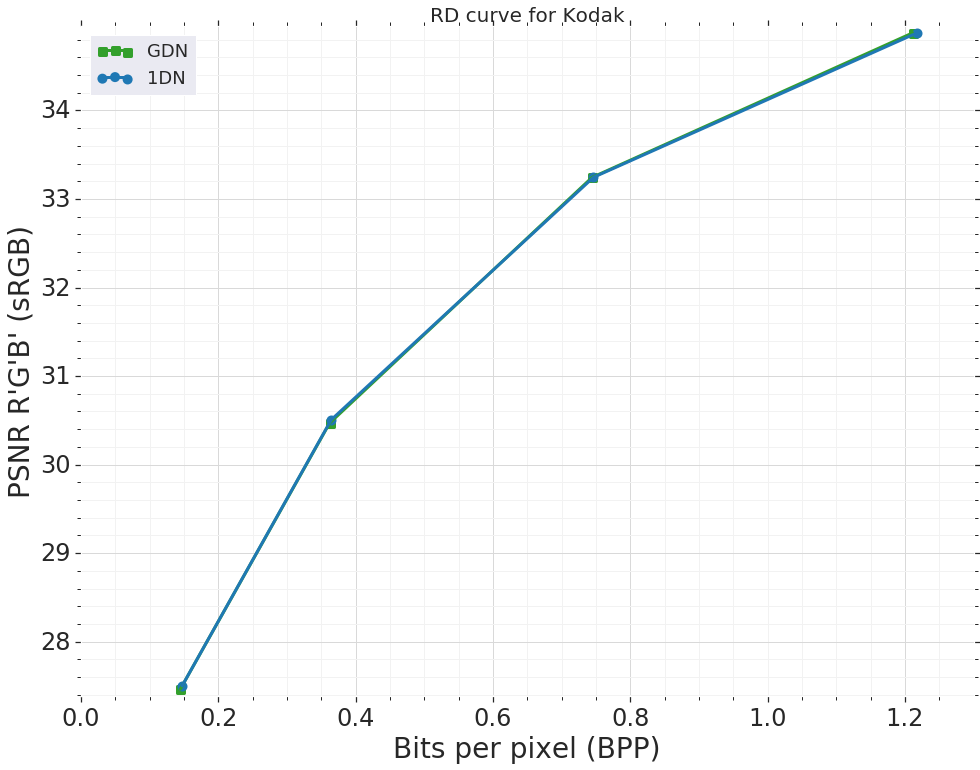}
  \caption{Rate-distortion plots on Kodak in PSNR for both GDN (with square/square root components) and 1DN (those without square/square root components). The lines overlap for the entirety of the rate-distortion curve, showing no compression performance degradation despite a significant computational simplification.
  }
  \label{fig:gdn}
\end{figure}

Divisive normalization (DN)~\cite{CaHe12} is a multivariate nonlinearity designed to model the responses of sensory neurons in biological systems, and has been shown to play a role in efficient information processing. GDN was introduced in \cite{BaLaSi16} as a probabilistic flow model, generalizing models such as independent subspace analysis and sparse coding. As an activation function, it is defined as:
\begin{equation}
\label{equ:gdn}
z_i = \frac {x_i} {\bigl( \beta_i + \sum_j \gamma_{ij} \, |x_j|^{\alpha_{ij}} \bigr)^{\varepsilon_i}},
\end{equation}
where $x_{i}$ and $z_{i}$ denote the input and output vectors, respectively, $ \alpha_{ij}$, $\beta_{i}$, $\gamma_{ij}$, $\varepsilon_{i}$ represent trainable parameters, and $i$, $j$ represent channel indexes. It has been demonstrated that, compared to common pointwise activation functions such as relu or tanh, GDN yields substantially better performance at the same number of hidden units and comes with a negligible increase in number of model parameters when used inside an autoencoder-style image compression model~\cite{Ba18}. It is also used in our Mean-Scale baseline. In their work, some parameters were fixed for simplicity ($\alpha_{ij} \equiv 2$ and $\varepsilon_i \equiv 0.5$). Given that computing a square root can take substantially more cycles than basic arithmetic, we re-examined this choice. We found that fixing $\alpha_{ij} \equiv \varepsilon_i \equiv 1$ doesn't hurt performance, while removing the need to compute square roots, and simplifying the function to basic arithmetic:
\begin{equation}
z_i = \frac {x_i} {\beta_i + \sum_j \gamma_{ij} \, |x_j|}.
\end{equation}
Figure~\ref{fig:gdn} shows a direct comparison of the rate--distortion performance for both GDN with square/square root exponents and our basic arithmetic version on the Kodak image set \cite{kodak}. To make this comparison, we followed \cite{Ba18} in the choices for hyperparameters such as optimization method and network architecture.

\section{Training with Group Lasso Regularization}
\label{section:glr}
\label{section:training}

For a given architecture, the number of filters in each layer is often an easy to ignore hyper parameter. The  majority of CNN architecture either use a constant number of filters per layer, or a simple doubling scheme each time the spatial resolution is reduced. It is intuitive that these naive approaches do not necessarily provide the optimal structure for trading off task performance with computational cost\cite{efficientnet}. We use a simple adaptation of the MorphNet approach\cite{gordon2017morphnet} which uses a convex sparsity inducing regularizer in order to learn the number of filters in each layer. Some techniques use the batch-norm scaling factors to determine the relative activity of each channel. Since the Mean-Scale model doesn't use batch-norm, we use a weighted Group Lasso \cite{yuan2006model} regularization term. For a fully-connected layer with matrix $W$ and $n$ outputs this results in: 

\begin{equation}
FLOP(W) = \sum_{j=1}^n \frac{1}{\sqrt{D}}  \|W_j \|_2
\end{equation}
where $W_j$ stands for all the weights associated with the $j$'th output. This applies analogously to convolutions.

In our experiments we use the FLOP-regularizer with a threshold of $0.001$ to determine which output is active. We follow the MorphNet procedure, by first learning several structures by applying increasing regularization strength, and then after the structure convergence, we retrain the models given the learned structure.

Many of the architectures heavily constrained the mean prediction network in the hyper decoder. Manual tuning was required to ensure that the mean prediction network matched in tensor output shape to the scale prediction network, meaning that slightly more computation was manually added in to make valid architectures.

We do not store any checkpoints or reuse any weights between these two phases, the second phase of training is from scratch.

For the structure learning process, we use the code publicly available code from \cite{morphnet_oss}.

\begin{table}
    \begin{center}
    \scriptsize
        \begin{tabular}{c|c|c|c}
             \hline
             Hyper Decoder 1 & Decoder 1 & Hyper Decoder 2 & Decoder 2 \\
             \hline
             5x5deconv,2,76  & 5x5deconv,2,150 & 5x5deconv,2,10  & 5x5deconv,2,25 \\
             5x5deconv,2,107 & 5x5deconv,2,89  & 5x5deconv,2,10  & 5x5deconv,2,21 \\
             3x3deconv,1,320 & 5x5deconv,2,81  & 3x3deconv,1,320 & 5x5deconv,2,19 \\
                             & 5x5deconv,2,3   &                 & 5x5deconv,2,3 \\
        \end{tabular}
        \caption{Two example decoder structures learned using MorphNet. Hyper Decoder 1 and Decoder 1 are paired examples for a MSE optimized network at 16\% of the computation of the Larger Mean-Scale models while Hyper Decoder 2 and Decoder 2 are paired for a MS-SSIM optimized network at only 6.7\% of the computation. We use the notation of CxC[conv or deconv],S,F to describe either a convolutional or deconvolutional layer with a kernel of CxC, a stride of S with F output channels.}
        \label{tab:examples}
    \end{center}
\end{table}

For training, we used the Larger Mean-Scale model architectures with a learning rate of 1e-4 using Adam as an optimizer with $\beta_1=0.9$, $\beta_2=0.999$ and $\epsilon=1e-8$. In addition to training for the usual rate-distortion term, the first pass of the training was modified to include the FLOP-regularizer on weights in the decoder and hyper-decoder, represented by $FLOP(W)$. A sweep of weighting terms for alpha was done in parallel with the usual sweep on lambda terms.
\begin{equation}
L = D*\lambda + R + FLOP(W)*\alpha
\end{equation}

The sweeps over $\lambda$\ depend on the distortion metric used:
for MSE we used $\lambda \in \{0.1\cdot 2^{t-6}\}_{t=0}^7$ and 
for MS-SSIM we used $\lambda \in \{2^{t-1}\}_{t=0}^7$.
For both distortion functions, the $\alpha$\ sweep was over 
$\alpha \in \{0.001 \cdot 0.2^{t}\}_{t=0}^{11}$.


As in \cite{NIPS2018_8275} we do the $\lambda$\ sweeps over both MSE and MS-SSIM~\cite{wang2003multiscale} as a distortion. MS-SSIM is reported in a logarithmic space to better visualize results in the ranges closer to loseless. These models were trained for roughly 100,000 steps in the first phase. The sweeps across $\lambda$\ and $\alpha$\ result in 207 unique decoder architectures created in the first phase of training.

In the second phase of training, we reduce the convolutional filters in the decoder down to values determined by the FLOP-regularizer and train for over $1e6$ steps or to convergence. All of our models are trained using the TensorFlow~\cite{tensorflow2015-whitepaper} library.

\section{Results}
\label{section:results}
The training for FLOP optimization created 207 unique decode architectures. These optimized networks shared a few interesting traits that are worth noticing here.

None of the optimized networked shrank the scale prediction network in the hyper decoder. 
This could be explained by the fact that this part of the model only makes up for a fraction of the overall compute, hence little is gained by shrinking it.

The FLOP-regularized network tended to shrink most in the second and third decoder layers. This is certainly due to the fact that the image is close to the original size spatially and at this point is either one-half or one-quarter in both dimensions and a large amount of compute is needed due to the fully convolutional nature of the network and that all pixels must be processed.

A full listing of all architectures, along with the final loss function and $\lambda$\ values are available in the Supplementary Materials: \url{https://storage.googleapis.com/compression-ml/cenic/Supplementary_Materials_Computationally_Efficient_Neural_Image_Compression.pdf}.

\subsection{Inference Performance}
\label{section:performance}
The run-time information was gathered on a single PC with a Intel Xeon W-2135 CPU @ 3.70GHz (no GPU or other accelerators are leveraged). A sweep of models described in Section~\ref{section:training} were ran on all images in the Kodak~\cite{kodak} dataset. We focus on decode times of the networks in our performance evaluation, this includes not only the neural network decoding but the range coder implementation to create an actual bitstream.

\begin{figure}[tb]
  \centering
  \includegraphics[width=0.90\linewidth]{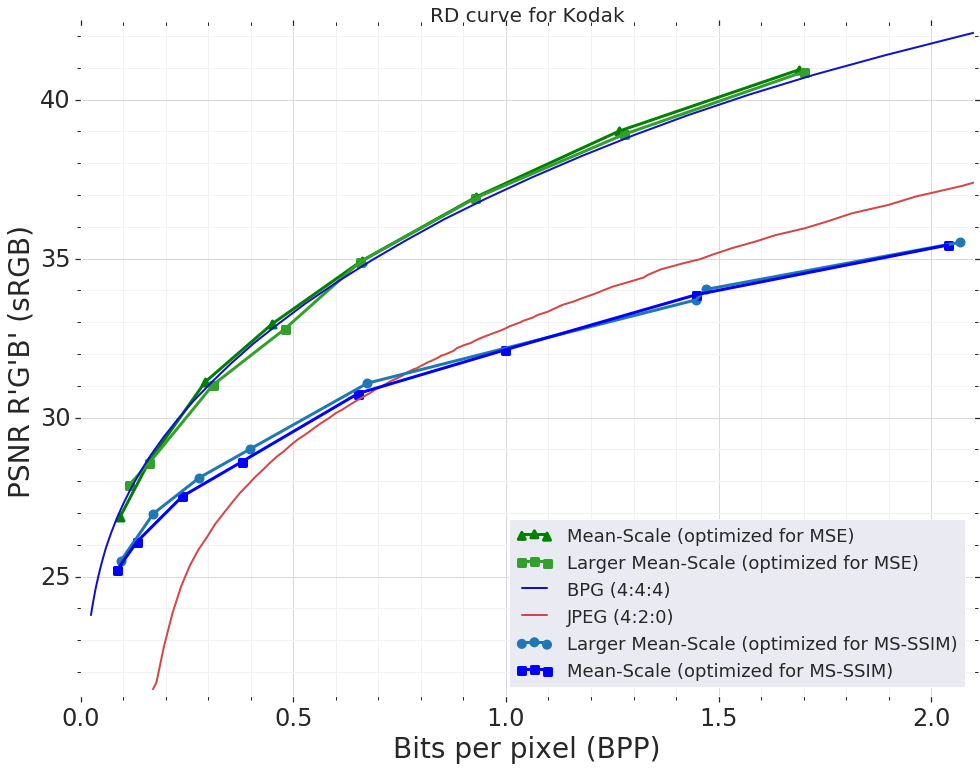}
  \includegraphics[width=0.90\linewidth]{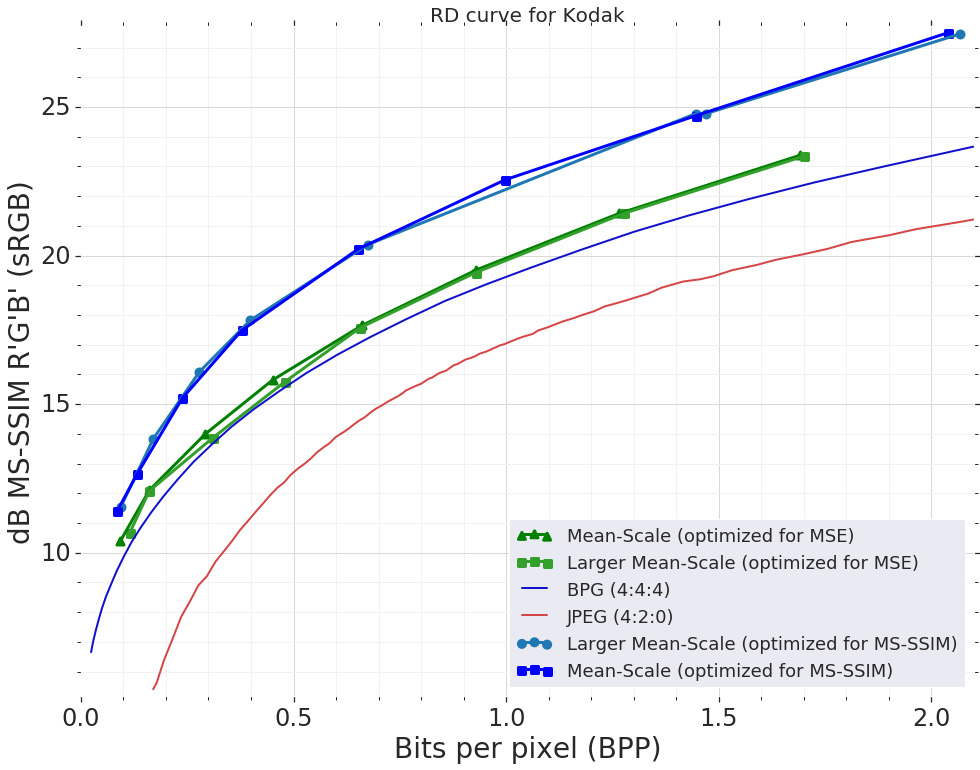}
  \caption{Rate-distortion plots on Kodak in PSNR (top) and dB MS-SSIM (bottom) for the Mean-Scale and Larger Mean-Scale networks. This is to show the Mean-Scale models are near that of what was presented in \cite{NIPS2018_8275} and that blindly adding additional capacity to the Larger Mean-Scale models do not produce significantly improved results. dB MS-SSIM is calculated as $-10*log_{10}(\text{MS-SSIM})$.}
  \label{fig:rd_baseline}
\end{figure}

We compare the Mean-Scale and Larger Mean-Scale performances on Kodak in Figure~\ref{fig:rd_baseline}. This is to show that that greatly increased capacity doesn't directly yield significantly different performance on the rate-distortion curve, but does significantly increase run-time and training time.

\begin{figure}[tb]
  \centering
  \includegraphics[width=0.90\linewidth]{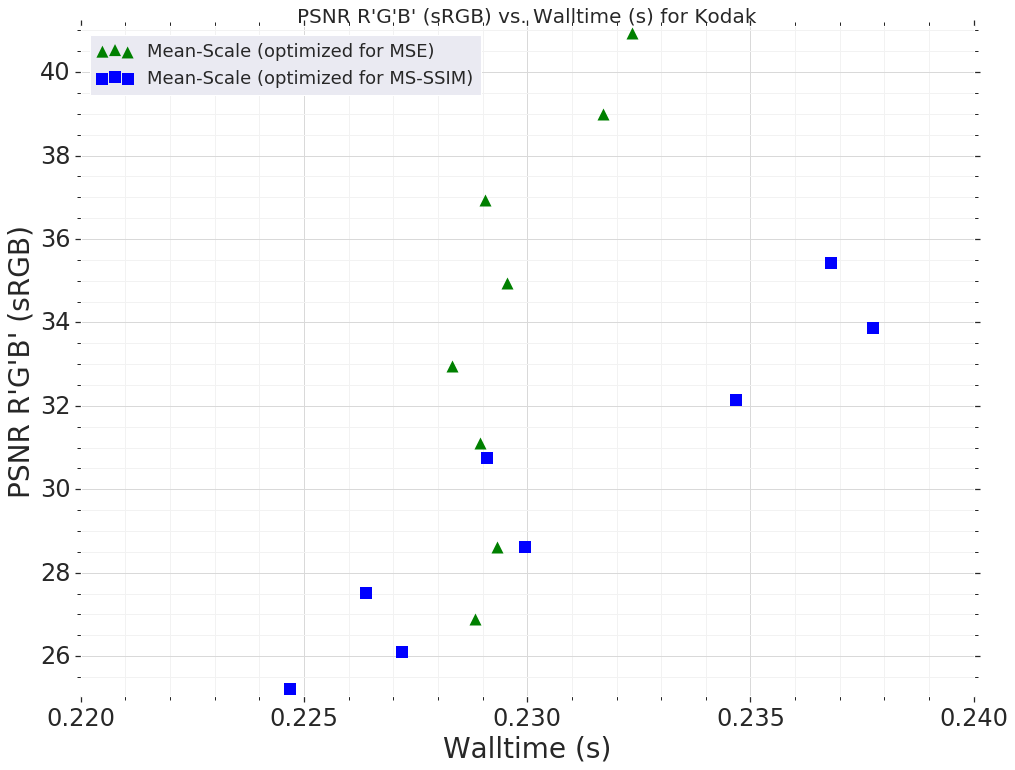}
  \includegraphics[width=0.90\linewidth]{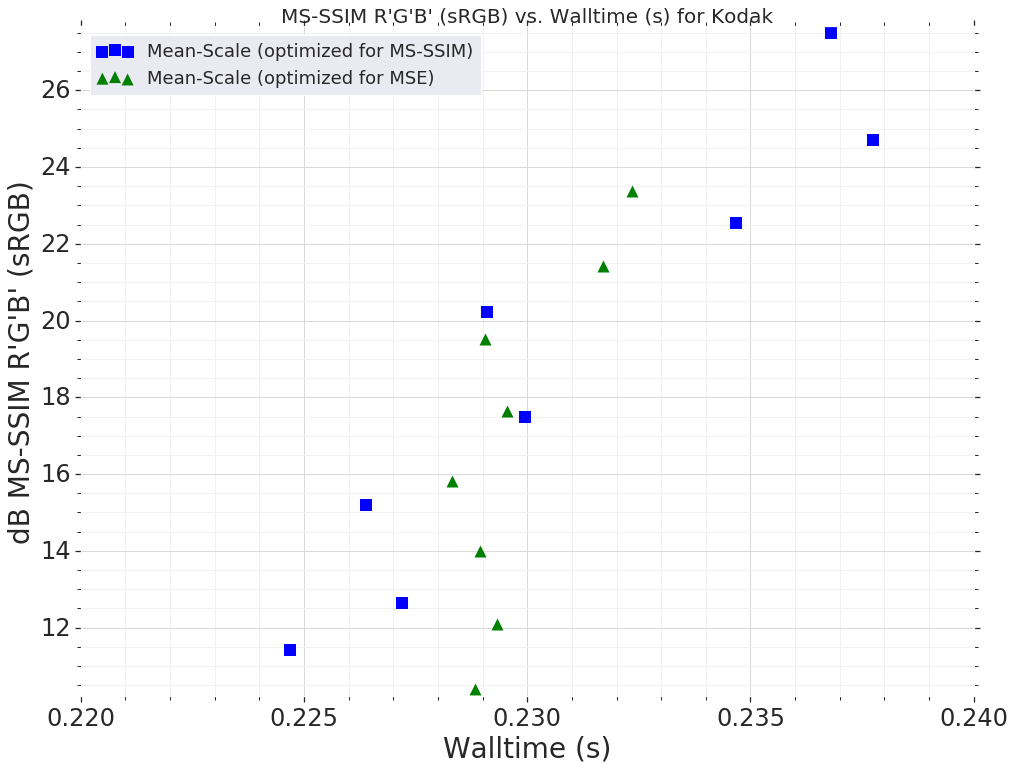}
  \caption{Run-time--distortion plots on Mean-Scale model for Kodak in PSNR (top) and dB MS-SSIM (bottom) on an Intel CPU. dB MS-SSIM is calculated as $-10*log_{10}(\text{MS-SSIM})$.
  }
  \label{fig:runtime_baseline}
\end{figure}

We also include a run-time--distortion graph for the Mean-Scale model in Figure~\ref{fig:runtime_baseline}. It's worth noticing that the variance for the MSE networks is lower than the MS-SSIM optimized networks. Another interesting point of analysis is that the run-time for the MS-SSIM network is almost 10\% slower on the higher quality end. This is a concrete example of the loss function mattering, as it has a concrete effect on run-time despite all of the points in the plot having the exact same network structure.

The run-time difference is due to the complex interactions between reconstruction quality, how the images are encoded and the probability model, e.g. a model that produces highly probable codes and distributions with many unused symbols decode faster in a range coder implementation than one where the codes are evenly distributed. It is also worth noting that the majority of the time spent is within the neural network's hyper-decoder and decoder in reconstructing the intermediate and final images from the decoded bitstream.

\subsection{Performance on GDN activations}
As shown in \cite{Ba18} and \cite{balle2018integer}, GDN is a powerful activation function for neural image compression networks. TensorFlow operation
level profiling shows the amount of time spent in average on each of the three GDN activations in the decoder layer. The change to the parameters  $\alpha_{ij}$ and $\varepsilon_i$ removes a square and a square root operation from the model. Table ~\ref{tab:gdn_savings} shows the relative and absolute savings within GDN for this simplification. Overall, this results in a 14.33 ms reduction (21.4\% savings) within the GDN operation and a 2.7\% overall speed-up in the Mean-Scale models.

\begin{table}
    \begin{center}
    \scriptsize
        \begin{tabular}{c|c|c|c}
             \hline
              & 1st Layer & 2nd Layer & 3rd Layer \\
             \hline
             Square & 0.327 ms (11.6\%) & 1.705 ms (12.9\%) & 6.843 ms (13.4\%) \\
             Sqrt & 0.137 ms (4.9\%) & 0.998 ms (7.5\%) & 4.32 ms (8.5\%) \\
        \end{tabular}
        \caption{The absolute (and relative) time saved within the GDN activation after simplifying $\alpha_{ij}$ and $\varepsilon_i$ in Equation ~\ref{equ:gdn}. GDN is generally implemented with a the following TensorFlow ops: Square, Conv, BiasAdd, Sqrt, Mul. The Square and the Sqrt are simplified out in our faster formulation.}
        \label{tab:gdn_savings}
    \end{center}
\end{table}

\subsection{Performance on CENIC architectures}

\begin{figure}[tb]
  \centering
  \includegraphics[width=0.90\linewidth]{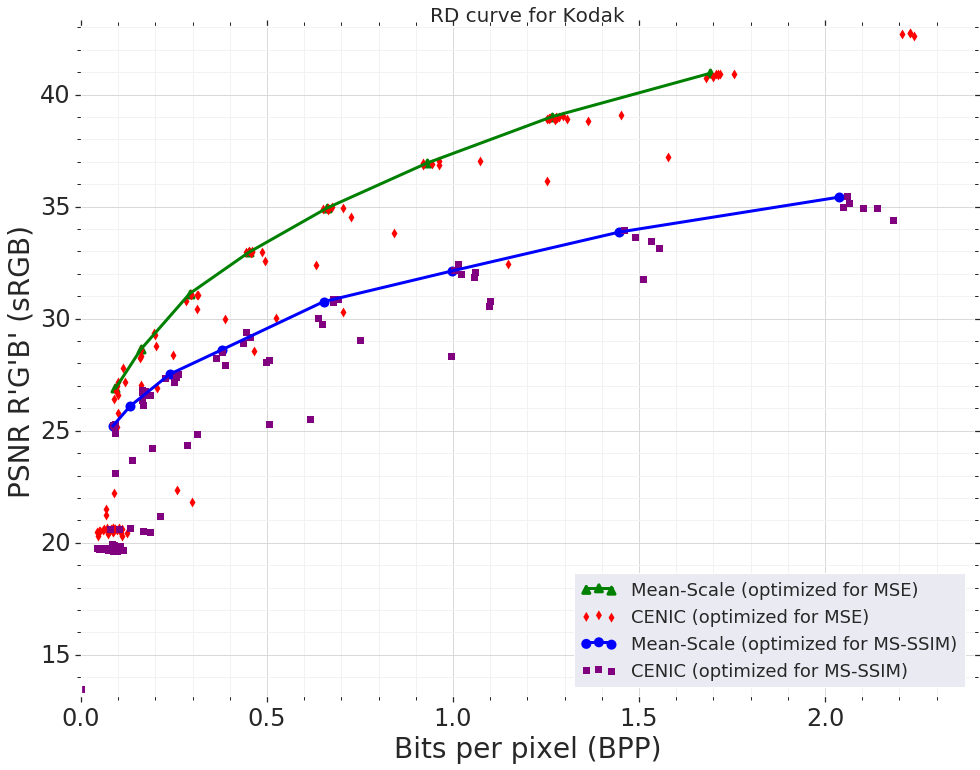}
  \includegraphics[width=0.90\linewidth]{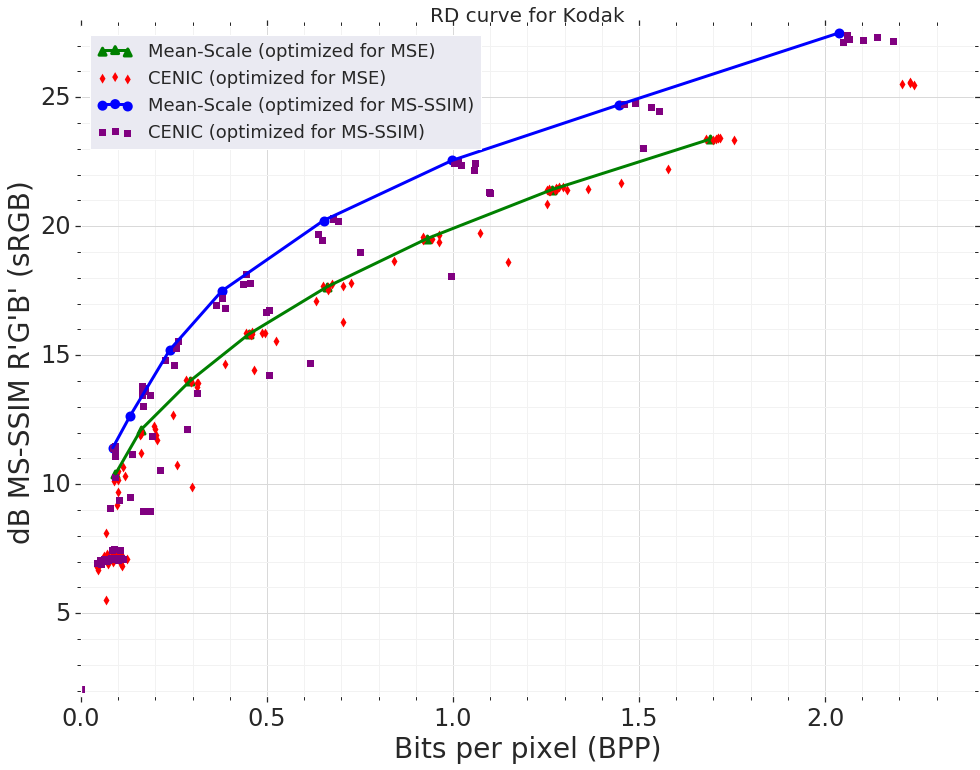}
  \caption{All learned architectures plotted with the Mean-Scale results with respect to MSE (top) and MS-SSIM (bottom). Due to the fact that we're doing an aggressive FLOP-regularized sweep over each rate point, we expect many low performance points to exist. dB MS-SSIM is calculated as $-10*log_{10}(\text{MS-SSIM})$.
  }
  \label{fig:allarchs}
\end{figure}

For the CENIC models, we graph the distortions compared to the Mean-Scale models with respect to average decoding time on Kodak. We draw a line connecting the various points for the Mean-Scale models so one can more easily see the frontier of faster models (on the left) vs. slower models (on the right).

\begin{figure}[tb]
  \centering
  \includegraphics[width=0.90\linewidth]{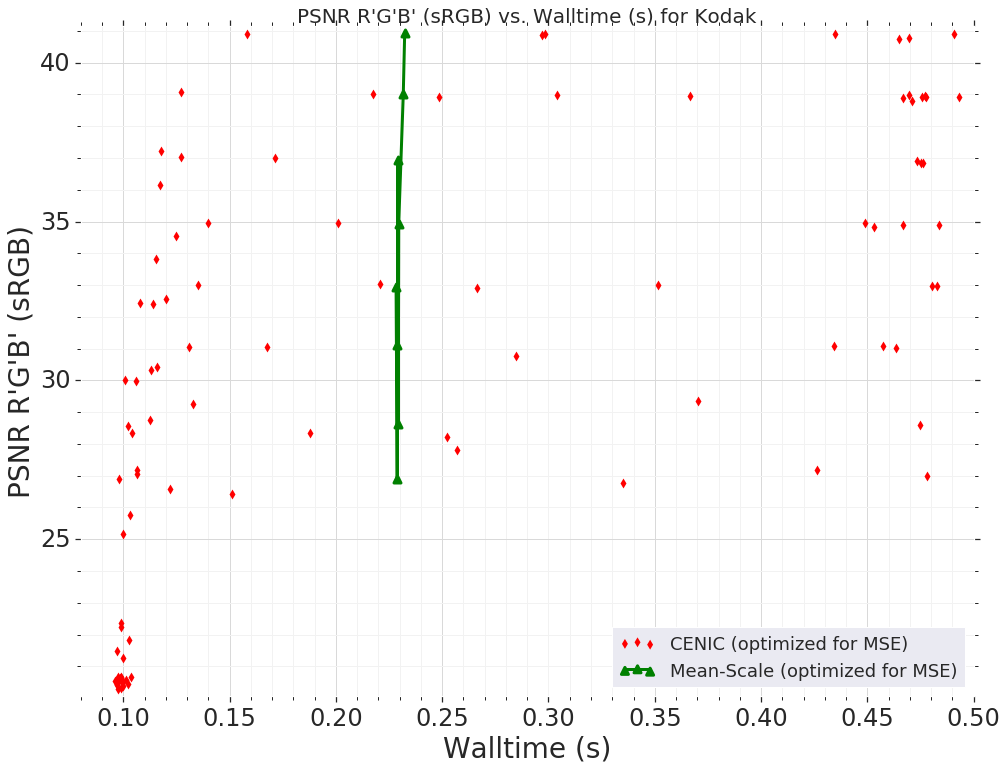}
  \includegraphics[width=0.90\linewidth]{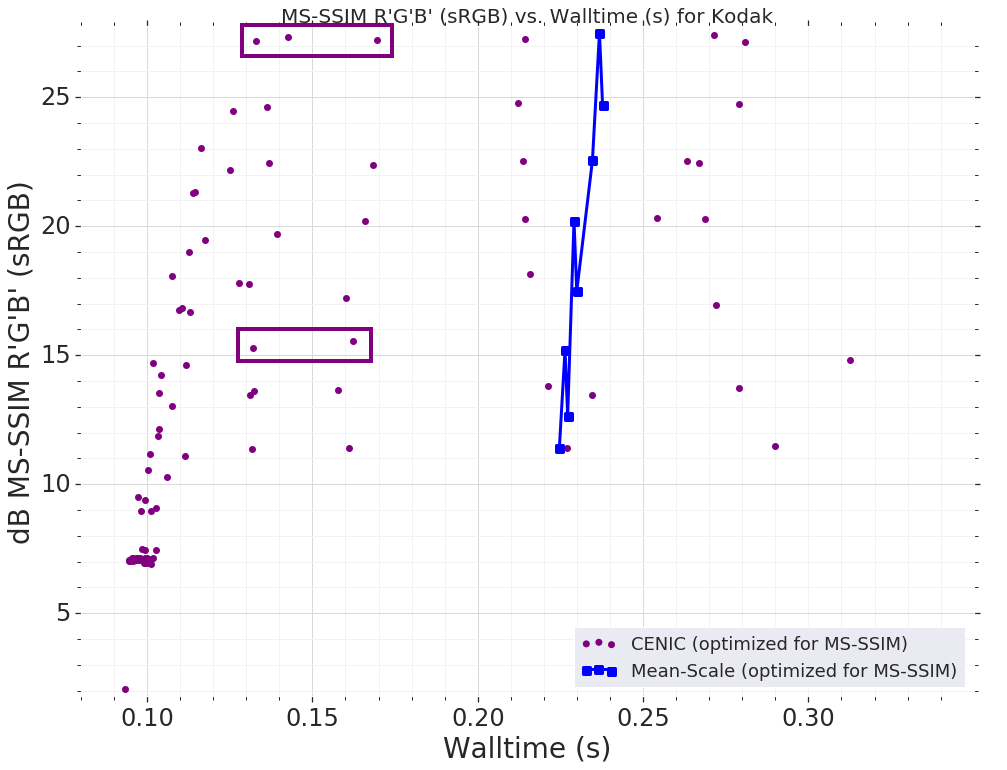}
  \caption{All learned architectures plotted with the Mean-Scale results models optimized on MSE (top) and dB MS-SSIM (bottom) vs. average decode time. dB MS-SSIM is calculated as $-10*log_{10}(\text{MS-SSIM})$.
  }
  \label{fig:cenicarchs}
\end{figure}

A number of patterns appear in these run-time--distortion plots. The first is that for MSE optimized networks, many of them appear to the right of the vertical baseline (around 0.23 seconds). This is because we use the Larger Mean-Scale baseline (with more capacity) to begin the regularization process and many of these networks were not able to be decreased back to the baseline model. However, more of the MS-SSIM optimized models appear left of the vertical baseline, indicating that it is easier to optimize for FLOP performance when also training for this loss function.

The second is that there are more points clustered in the lower-left corner of the graph for MS-SSIM vs MSE, meaning that low-quality (synonymous with low bitrate) models occurred for MS-SSIM vs MSE. These two graphs reinforce that taking into account the loss function and evaluation metric are key to developing fast, high-quality compression models.

\begin{figure}[tb]
  \centering
  \includegraphics[width=0.90\linewidth]{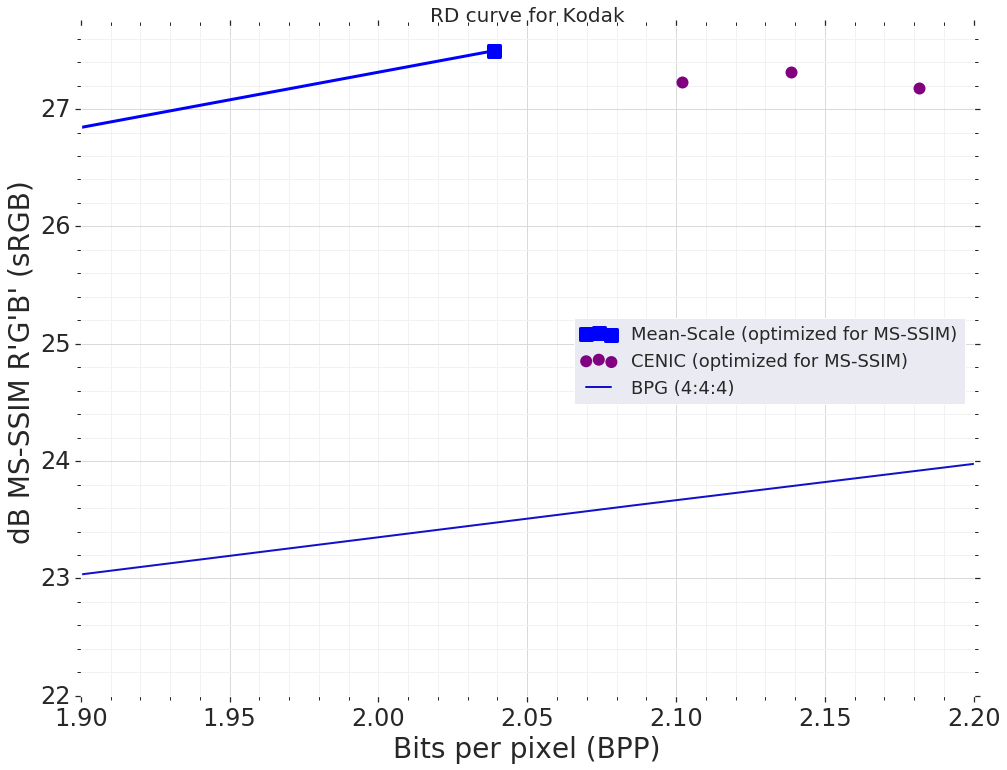}
  \includegraphics[width=0.90\linewidth]{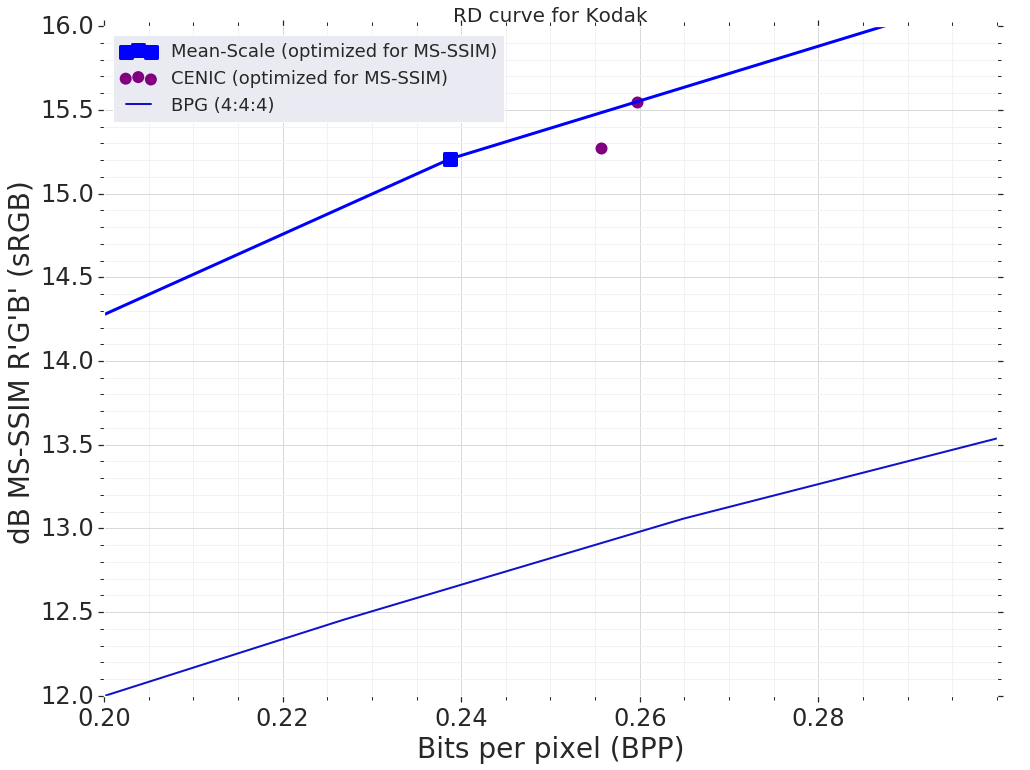}
  \caption{A closer look at CENIC models that operate at a similar distortion but much faster run-time than the Mean-Scale model. The higher quality models for comparison are in the top figure while the mid quality models are at the bottom.
  dB MS-SSIM is calculated as $-10*log_{10}(\text{MS-SSIM})$.}
  \label{fig:cenichighend}
\end{figure}

Let's dive in on a few of the accelerated points we found in the MS-SSIM search. First the higher quality points (grouped in the top box of Figure~\ref{fig:cenicarchs}b). These three models (architectures 5, 2 and 10 in the Supplementary Materials) have slightly decreased rate--distortion performance but is still much higher than the BPG baseline as shown in Figure~\ref{fig:cenichighend} and their architectures can be compared in Table~\ref{tab:cenichighendtab}.

\begin{table}
    \begin{center}
    \scriptsize
        \begin{tabular}{c|c||c|c}
             \hline
             Hyper Decoder 5 & Decoder 5 & Hyper Decoder 2 & Decoder 2  \\
             \hline
             5x5deconv,2,76  & 5x5conv,2,79 & 5x5deconv,2,40  & 5x5deconv,2,149  \\
             5x5deconv,2,107 & 5x5conv,2,22 & 5x5deconv,2,67  & 5x5deconv,2,35  \\
             3x3deconv,1,320 & 5x5conv,2,43 & 3x3deconv,1,320 & 5x5deconv,2,39  \\
                             & 5x5conv,2,3  &                 & 5x5deconv,2,3   \\
            \hline
            \hline
            Hyper Decoder 10 & Decoder 10 & & \\
            \hline
             5x5deconv,2,66 & 5x5deconv,2,180 & & \\
             5x5deconv,2,95 & 5x5deconv,2,58 & & \\
             3x3deconv,1,320 & 5x5deconv2,73 & &\\
                              & 5x5deconv2 & &\\
        \end{tabular}
        \caption{The three high-quality MS-SSIM optimized CENIC models representing the scatter plot in Figure~\ref{fig:cenichighend}a. We use the notation of CxC[conv or deconv],S,F to describe either a convolutional or deconvolutional layer with a kernel of CxC, a stride of S with F output channels.}
        \label{tab:cenichighendtab}
    \end{center}
\end{table}

If we look at two models in the mid quality range of MS-SSIM (CENIC architectures 32 and 37), we find two models with substantial decoder speed-ups while performing very close to the Mean-Scale network shown in the bottom half of Figure~\ref{fig:cenichighend} and their architectures are shown in Table~\ref{tab:cenicmidarchs}. All architecture descriptions and runtimes are provided in the Supplementary Materials.

\begin{table}
    \begin{center}
    \scriptsize
        \begin{tabular}{c|c||c|c}
             \hline
             Hyper Decoder 32 & Decoder 32 & Hyper Decoder 37 & Decoder 37  \\
             \hline
             5x5deconv,2,246  & 5x5conv,2,100 & 5x5deconv,2,110  & 5x5deconv,2,52  \\
             5x5deconv,2,170 & 5x5conv,2,126 & 5x5deconv,2,91  & 5x5deconv,2,99  \\
             3x3deconv,1,320 & 5x5conv,2,52 & 3x3deconv,1,320 & 5x5deconv,2,14  \\
                             & 5x5conv,2,3  &                 & 5x5deconv,2,3   \\
        \end{tabular}
        \caption{The two mid-quality MS-SSIM optimized CENIC models representing the scatter plot in Figure ~\ref{fig:cenichighend}b. We use the notation of CxC[conv or deconv],S,F to describe either a convolutional or deconvolutional layer with a kernel of CxC, a stride of S with F output channels.}
        \label{tab:cenicmidarchs}
    \end{center}
\end{table}

\subsection{Performance differences between MS-SSIM and PSNR}
The clustering of models in Figure ~\ref{fig:allarchs} shows that especially at the high bitrates of neural image compression, MS-SSIM models suffer less than MSE models. This can be seen by the much larger dips in performance above 1 bit-per-pixel. Below 1 bit-per-pixel the MS-SSIM models lose substantially more performance and PSNR at similar capacities.

This observation is in line with anecdotal evidence of MS-SSIM being easier to train at higher bitrates and MSE being easier to train at lower bitrates--the loss functions and the capacity of the model play a key role of interacting together. It also explains the tendency of recent end-to-end learned compression research to report results in MS-SSIM, as it is easier to get high performance results on an unoptimized architecture.

\subsection{Performance differences between CPU and GPU}
Though we are looking at the decoding characteristics from a general purpose CPU perspective, it is worth understanding how Mean-Scale network's decoding time varies between CPU and GPU platforms. The GPU used for these numbers is a NVIDIA TitanXp on the a system with the same  CPU as discussed above in Section~\ref{section:results}.

\begin{figure}[tb]
  \centering
  \includegraphics[width=0.90\linewidth]{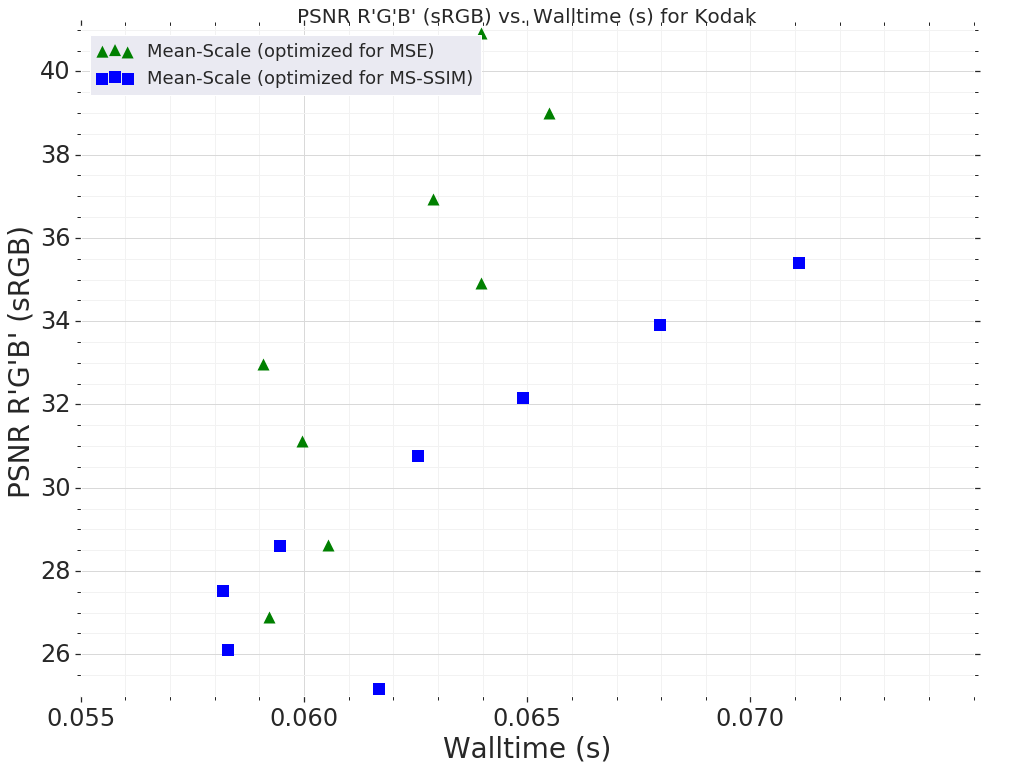}
  \includegraphics[width=0.90\linewidth]{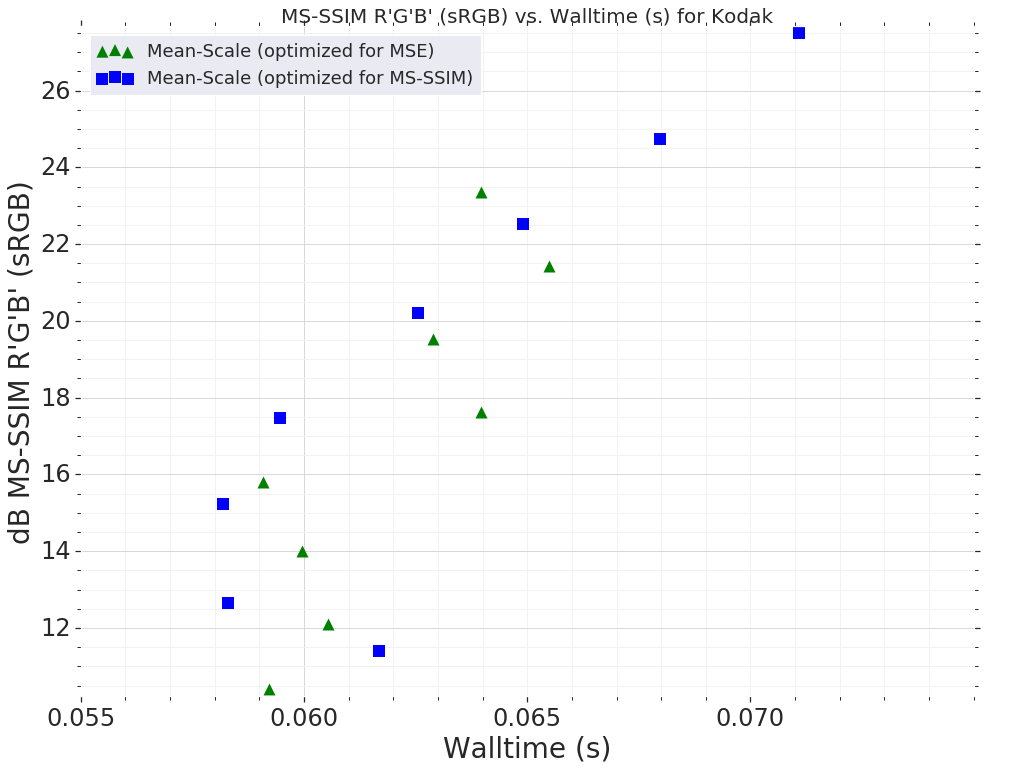}
  \caption{Run-time--distortion plots on Mean-Scale model for Kodak in PSNR (top) and dB MS-SSIM (bottom) on a NVIDIA TitanXp GPU. dB MS-SSIM is calculated as $-10*log_{10}(\text{MS-SSIM})$.}
  \label{fig:runtime_gpu}
\end{figure}

From Figure~\ref{fig:runtime_gpu}, it is clear that, as expected, the Mean-Scale model runs much faster on GPU. With the average CPU run-time being around 230 milliseconds and the average GPU run-time being around 62 milliseconds (a 3.7x speed-up). 

\section{Conclusions}
\label{section:conclusions}
This work on analyzing the run-time and performance of neural image compression networks has led to a few of the following conclusions.

First, in the process of exploring FLOP-regularized architectures, we saw that simply adding more capacity to each layer led to no additional rate-distortion gains. This leads us to the conclusion that these four layer decoder and three layer hyper-decoder networks have already been well tuned for high rate-distortion performance, especially at the high rate/high quality end of the spectrum and this is done at significant costs to run-time and train time.

Second, we found that automating architecture search while constraining a network to a particular run-time and the use of methods like MorphNet that allows optimization directly on FLOPs is important because the hyperparameter search space is too large to effectively find good architectures manually.

The CPU and GPU comparisons of the models show that the compute platform and implementation matter. Running on a powerful GPU yielded almost a 3x improvement in run-time. Optimizing for MSE also produced lower variance decoding time across both platforms. The variance is due to the complex interactions between the neural network coding and probability distributions affecting encode and decode time of the range coder, since the amount of computation in the neural network components were fixed for these models.

After looking at the GDN activation functions, we also recommend the simplification to the $\alpha_{ij}$ and $\varepsilon_i$ parameters to reduce decoding time while keeping rate--distortion performance nearly identical.

Lastly, MS-SSIM optimized models were able to be optimized for compute easier than their MSE optimized counterparts. This is encouraging because MS-SSIM generally correlates better with human perceptual quality than MSE~\cite{wang2003multiscale}. Optimizing for decoder speed not only allows for more practical, faster inference models, but also allows for faster neural network training due to having less parameters and fewer gradients.

\section{Future Work}
If working in real-time applications, it is clear that an arbitrarily slow encoder is not feasible. For example, when a photo is taken on mobile, it is generally expected to be viewable quickly, which means a reasonable encoding time is needed. Using MorphNet, both encoder and decoder networks can be optimized jointly to discover new architectures that balance real-time needs with compression performance.

Neural image compression networks that operate fully convolutionally also tend to produce large intermediate tensors, creating heavy stress on caches and internal memory systems. While we investigate the run-time components of these decoder networks, we did not look at the effects on the memory systems (though decreasing activations in these networks does strictly decrease the amount of memory required). These memory requirements may be more stringent in embedded applications or on special purpose accelerators where both memory and memory bandwidth are limited.

Finally, researching more perceptually-based image metrics that are differentiable (or approximately differentiable for the purposes of training neural networks) could find more optimal run-time--distortion trade offs.

{\small
\bibliographystyle{ieee_fullname}
\bibliography{egbib}
}

\end{document}


\title{Supplementary Materials for Computationally Efficient Neural Image Compression}

\author{Nick Johnston\\
Google Research\\
{\tt\small nickj@google.com}
\and
Elad Eban\\ {\tt\small elade@google.com}
\and
Ariel Gordon\\ {\tt\small gariel@google.com}
\and 
Johannes Ballé \\ {\tt\small jballe@google.com}
}

\maketitle

\section{List of Architectures}
We include a listing of all the architectures that were generated in our FLOP-regularized network. Additionally,
we have provided the decode runtimes for each Kodak image in the dataset. The Hyper Decoder and Decoder networks are described as in the paper with the notation of CxC[conv or deconv],S,F to describe either a convolutional or deconvolutional layer with a kernel of CxC, a stride of S with F output channels.
\onecolumn
\begin{longtable}{|l|l|l|l|l||l|}
\hline
\textbf{Architecture} & \textbf{Loss Function} & \textbf{Lambda} & \textbf{Hyper Decoder} & \textbf{Decoder} & \textbf{Kodak Decode Time} \\
\hline

\hline1 & MS-SSIM & 64.0 & 5x5deconv,2,10 & 5x5deconv,2,2 & Kodak 1: 0.093272\\ &  &  & 5x5deconv,2,10 & 5x5deconv,2,8 & Kodak 2: 0.100056\\ &  &  & 5x5deconv,2,320 & 5x5deconv,2,1 & Kodak 3: 0.098977\\ &  &  &  & 5x5deconv,2,3 & Kodak 4: 0.099448\\ &  &  &  &  & Kodak 5: 0.101219\\ &  &  &  &  & Kodak 6: 0.102863\\ &  &  &  &  & Kodak 7: 0.100162\\
\hline

\end{longtable}
\twocolumn